\documentclass[11pt,a4paper]{elsart}
\usepackage[english]{babel}
\usepackage{epsfig}
\usepackage{amsmath}
\usepackage{amsfonts}
\usepackage{amsbsy}
\usepackage{amssymb}
\usepackage{latexsym}
\usepackage{pennames}
\def\units#1{\hbox{$\,{\rm #1}$}}
\def\degrees{\hbox{${}^\circ$}}

\def\Journal#1#2#3#4{{#1} {\bf #2}, #3 (#4)}
 
\def\etal{{\it et\ al.}}

\def\NIMA{{\em Nucl. Instrum. Methods} A}

\def\PRD{{\em Phys. Rev.} D}

\def\ASP{{\em Astrop. Phys.}}
\def\APJ{{\em Astroph. J.}}

\begin{document}

\begin{frontmatter}

\title{Measurement of the residual energy of
muons in the Gran Sasso underground Laboratories} 

{\rm The MACRO Collaboration} \\
\nobreak\bigskip\nobreak
\pretolerance=10000
\author{M.~Ambrosio$^{12}$}, 
\author{R.~Antolini$^{7}$}, 
\author{A.~Baldini$^{13}$}, 
\author{G.~C.~Barbarino$^{12}$}, 
\author{B.~C.~Barish$^{4}$}, 
\author{G.~Battistoni$^{6,b}$}, 
\author{Y.~Becherini$^{2}$},
\author{R.~Bellotti$^{1}$}, 
\author{C.~Bemporad$^{13}$}, 
\author{P.~Bernardini$^{10}$}
% ...
\\
\vspace{1.5 cm}
\footnotesize
1. Dipartimento di Fisica dell'Universit\`a  di Bari and INFN, 70126 Bari, Italy \\
2. Dipartimento di Fisica dell'Universit\`a  di Bologna and INFN, 40126 Bologna, Italy \\
3. Physics Department, Boston University, Boston, MA 02215, USA \\
4. California Institute of Technology, Pasadena, CA 91125, USA \\
5. Department of Physics, Drexel University, Philadelphia, PA 19104, USA \\
6. Laboratori Nazionali di Frascati dell'INFN, 00044 Frascati (Roma), Italy \\
7. Laboratori Nazionali del Gran Sasso dell'INFN, 67010 Assergi (L'Aquila), Italy \\
8. Depts. of Physics and of Astronomy, Indiana University, Bloomington, IN 47405, USA \\
9. Dipartimento di Fisica dell'Universit\`a  dell'Aquila and INFN, 67100 L'Aquila, Italy\\
10. Dipartimento di Fisica dell'Universit\`a  di Lecce and INFN, 73100 Lecce, Italy \\
11. Department of Physics, University of Michigan, Ann Arbor, MI 48109, USA \\
12. Dipartimento di Fisica dell'Universit\`a  di Napoli and INFN, 80125 Napoli, Italy \\
13. Dipartimento di Fisica dell'Universit\`a  di Pisa and INFN, 56010 Pisa, Italy \\
14. Dipartimento di Fisica dell'Universit\`a  di Roma "La Sapienza" and INFN, 00185 Roma, Italy \\
15. Physics Department, Texas A\&M University, College Station, TX 77843, USA \\
16. Dipartimento di Fisica Sperimentale dell'Universit\`a  di Torino and INFN, 10125 Torino, Italy \\
17. L.P.T.P, Faculty of Sciences, University Mohamed I, B.P. 524 Oujda, Morocco \\
18. Dipartimento di Fisica dell'Universit\`a  di Roma Tre and INFN Sezione Roma Tre, 00146 Roma, Italy \\
$a$ Also Universit\`a  della Basilicata, 85100 Potenza, Italy \\
$b$ Also INFN Milano, 20133 Milano, Italy \\
$c$ Also Istituto IASF/CNR, 40129 Bologna, Italy \\
$d$ Also Universit\`a  di Trieste and INFN, 34100 Trieste, Italy \\
$e$ Also Dipartimento di Energetica, Universit\`a  di Roma, 00185 Roma, Italy \\
$g$ Also Institute for Space Sciences, 76900 Bucharest, Romania \\
$h$ Macalester College, Dept. of Physics and Astr., St. Paul, MN 55105 \\
$i$ Also Department of Physics, Pennsylvania State University, University Park, PA 16801, USA \\
$l $Also Dipartimento di Fisica dell'Universit\`a  della Calabria, Rende (Cosenza), Italy \\
$o$ Also Dipartimento di Scienze e Tecnologie Avanzate, Universit\`a  del Piemonte Orientale, Alessandria, Italy \\
$p$ Also U. Minn. Duluth Physics Dept., Duluth, MN 55812 \\
$q$ Also Dept. of Physics, MIT, Cambridge, MA 02139 \\
$r$ Also Intervideo Inc., Torrance CA 90505 USA \\
$s$ Also Resonance Photonics, Markham, Ontario, Canada\\
\thanks[CA]{Corresponding authors: F.~Loparco, M.~N.~Mazziotta\\
Fax: +39 080 5442470; e-mail: loparco@ba.infn.it, mazziotta@ba.infn.it}

\begin{abstract}

The MACRO detector
was located in the Hall B of the Gran Sasso underground Laboratories
under an average rock overburden of $3700~hg/cm^{2}$.
A transition radiation detector 
composed of three identical modules, covering a total horizontal
area of $36 \units{m^{2}}$, was installed inside the 
empty upper part of the detector 
in order to measure the residual energy of muons. 
This paper presents the measurement of the 
residual energy of single
and double muons crossing the apparatus. 
Our data show that double muons are more energetic than single 
ones. This measurement is performed over a standard 
rock depth range from $3000$ to $6500 \units{hg/cm^{2}}$.
 
\end{abstract}

\end{frontmatter}


\section{Introduction}

Underground muons are the remnants of the air showers initiated by the
collisions of primary cosmic rays with air nuclei. 
These secondary cosmic ray muons can
easily cross large amounts of matter and penetrate into
underground laboratories.   
The energy spectra of underground muons
depend on the energy spectra and composition 
of primary cosmic rays, on their interactions with air nuclei 
and on the muon energy loss in the rock.

In this paper the measurement of the 
energy spectra of underground single and double muons,
performed with a $36 \units{m^{2}}$ Transition Radiation 
Detector (TRD), that was installed in the empty upper part
of the MACRO detector~\cite{macro}, is presented. 
An analysis using part of the single 
muon data sample was already published~\cite{trd3}.
The final sample is approximately 10 times larger
and a correct evaluation of the systematics has been performed,
thus allowing us to make more reliable conclusions about
the spectra and the average residual energy of
single and double muon events.  
The TRD sub-detector, described in sec.~\ref{sec:trd}, 
uses for triggering purpose and for the measurement of the event 
multiplicity the larger area ($\sim 1000\units{m^2}$) of 
the streamer tube and scintillation counters systems. 
The data selection is described in sec.~\ref{sec:data}.

To evaluate the muon energy spectra from the TRD
data, we used two complementary approaches
described in sec.~\ref{sec:spectra}. In the first case, 
an unfolding procedure was applied, while 
in the second case a parameterization for the 
underground muon energy spectrum
was assumed and a best fit of the spectral parameters was
carried out. 

To study how the energy spectra of underground muons
are related to the primary cosmic ray 
spectrum and composition, a dedicated 
Monte Carlo simulation, described in sec.~\ref{sec:MC},
was performed. 
The results of the comparisons between data and Monte Carlo 
are discussed in sec.~\ref{sec:discussion}.

 
\section{Underground cosmic ray muons}
\label{sec:undmuon}

The ``all-particle'' flux of the primary cosmic 
radiation can be described by an inverse power law energy 
spectrum \cite{gais}, with differential flux given by:
\begin{equation}
\frac{dN}{dE} \propto E^{-(\gamma+1)}
\end{equation}
where $\gamma \approx 1.7$ for $E \leq 10^3 \units{TeV}$, 
$\gamma \approx 2.0$ 
for $10^3 \units{TeV} \leq E \leq 10^6 \units{TeV}$ 
and $\gamma \approx 1.5$ for $E \geq 10^6 \units{TeV}$.

From the primary spectrum, it is possible to evaluate the 
energy spectrum at the Earth surface of secondary 
uncorrelated muons, which 
is given, with good approximation, by \cite{gais}:
\begin{equation}
\label{eq:musurf}
\frac{dN_{\mu}}{d \mathcal{E} d\Omega} 
\approx
const \cdot \mathcal{E} ^{-(\gamma+1)} \cdot 
\left(
\frac{1}{1 + a \mathcal{E}  \cos \theta}   +
\frac{0.054}{1 + b \mathcal{E}  \cos \theta} 
\right)
\end{equation}
where $\mathcal{E}$ is the muon energy at the surface, 
$a=1.1/115\units{GeV}$ and $b=1.1/850\units{GeV}$. 
The first and the second term in parenthesis of 
equation~\ref{eq:musurf} represent
the contributions of muons from $\Pgp$ and $\PK$ decays,
respectively. 
In the limit of high energies, an approximate expression of the muon
surface energy spectrum has the simple form:
\begin{equation}
\label{eq:musurf1}
\frac{dN_\mu}{d\mathcal{E}} = const \cdot \mathcal{E}^{-\alpha}
\end{equation}
where $\alpha = \gamma + 2 \approx 3.7$. The surface muon spectral 
index is therefore related to the primary cosmic ray spectral index.

The energy spectrum of underground muons can be derived taking into
account the process of energy loss in the rock, which is 
assumed to have the form: 
\begin{equation}
\label{eq:dedx}
\frac{dE}{dh} = - \left( \lambda + \beta E \right)
\end{equation}
where $dh$ is a thin rock slab (usually in $\units{g~cm^{-2}}$), 
$\lambda$ is the contribution from the ionization energy loss 
and $\beta E$ is the contribution from radiative processes 
(bremsstrahlung, pair production and muon hadroproduction). 
The parameters $\lambda$ and $\beta$ are functions of the muon energy, 
but for practical purposes can be assumed as constants~\cite{gais}.
The quantity $\epsilon = \lambda / \beta$ is called the critical energy 
and is defined as the energy value above which the radiative processes 
become dominant. 

With the above assumptions, the general solution of 
equation \ref{eq:dedx} is:
\begin{equation}
\label{eq:muund}
E_{\mu} = \left(\mathcal{E}  + \epsilon \right) e^{-\beta h} -
\epsilon.
\end{equation} 
where $E_{\mu}$ is the muon energy after crossing the rock 
slant depth $h~(g~cm^{-2})$. 
The underground muon energy spectrum can be thus obtained 
from equations \ref{eq:musurf1} and \ref{eq:muund} 
using the following relation:
\begin{equation}
\frac{dN}{dE_{\mu}} = 
\left[ \frac{dN}{d\mathcal{E} } \right]_{\mathcal{E} = \mathcal{E} (E_{\mu})} 
\cdot \frac{d\mathcal{E} }{dE_{\mu}} 
\end{equation}
and is given by: 
\begin{equation}
\label{eq:undspec}
\frac{dN}{dE_{\mu}} = const \cdot \left( E_{\mu} + \epsilon 
(1 - e^{-\beta h}) \right)^{-\alpha}.
\end{equation} 

From eq.~\ref{eq:undspec}, the average underground muon 
energy at depth $h$ is:
\begin{equation}
 \label{eq:emean}
 \langle E_{\mu} \rangle = 
 \frac{\epsilon (1 - e^{-\beta h} )}{\alpha - 2}
\end{equation}
and its asymptotic value is $\epsilon / (\alpha - 2)$.
At great depths $h$, the underground muon 
energy spectrum given by eq.~\ref{eq:undspec} is almost flat 
for $E \ll \langle E_{\Pgm} \rangle$, and then decreases 
with energy.


\section{The MACRO TRD}
\label{sec:trd}

Transition Radiation (TR) is the process of the emission 
of X-ray photons occurring when an ultrarelativistic charged 
particle crosses the boundary between two materials with 
different dielectric constants.
The most important features of TR are that
the TR yield is roughly proportional to the Lorentz factor
$\gamma$ of the radiating particle over a wide range of $\gamma$,
and the emission probability of a TR X-ray is of the order 
of $\alpha \approx 1/137$.
If the rest mass $m_{0}$ of the radiating particle is known, 
a measurement of its Lorentz factor $\gamma$ also allows 
one to evaluate the energy as $E = m_{0}c^{2} \gamma$. 
TRDs can provide energy measurements over ranges typically 
spanning one order of magnitude. 

Due to the characteristic dependence of the TR yield on the 
Lorentz factor $\gamma$, TRDs were proposed~\cite{trd1,trd2} 
for the measurement of energies of underground cosmic ray muons in
the $\units{TeV}$ region. The TRD operated inside the 
MACRO~\cite{macro} detector in the Gran Sasso underground 
Laboratory (LNGS) collected data from April 1995 to June 2000. 
It was designed to be sensitive to the energy region 
between $100 \units{GeV}$ and $1 \units{TeV}$.
  
The detector consisted of three modules covering a total 
horizontal area of $\sim 36 \units{m^{2}}$. Each module
was composed by eleven $10 \units{cm}$ thick radiator
layers, interleaved by ten planes of $32$ proportional 
tubes $6 \units{m}$ long and with a square cross
section of $6 \times 6 \units{cm^{2}}$.
A detailed description of the detector is given in~\cite{trd3,trd2}.
The radiator material (Ethafoam 220) contains
cells of $\sim 35 \units{\mu m}$ wall thickness and
$\sim 900 \units{\mu m}$ spacing, ensuring a threshold
Lorentz factor $\gamma_{th} \approx 10^{3}$ and
a saturation Lorentz factor $\gamma_{sat} \approx 10^{4}$,
that correspond to the muon energy range between
$\sim 100 \units{GeV}$ and $\sim 1 \units{TeV}$. 

The proportional tubes were filled with an 
$Ar (90\%) - CO_{2} (10\%)$ gas mixture and were 
operated at a gain of $\sim 10^{3}$. They were
equipped with a cluster counting read-out electronics.
Wire signals were sharply differentiated and compared
to a threshold corresponding to an energy deposit
of $\sim 5 \units{keV}$. In this way it was possible
to discriminate $\delta$-ray background from X-ray 
photoelectrons producing pulses exceeding the threshold
amplitude. For each event the pulses with amplitude
greater than the threshold (``hits'') were counted
in all the proportional tubes.

The third TRD module was partially equipped with 
ten $1 \units{mm}$ thick aluminum foils 
of $2 \times 2 \units{m^{2}}$ area, that were inserted 
between each radiator and the tube plane below, 
in the terminal part of the module.
The aluminum foils absorbed the TR emitted
by muons in the upstream radiator layers.
Using this technique a sample of muons were collected
with only the ionization loss measured.

\begin{figure}
\begin{center}
\resizebox{13.0cm}{13.0cm}{\includegraphics{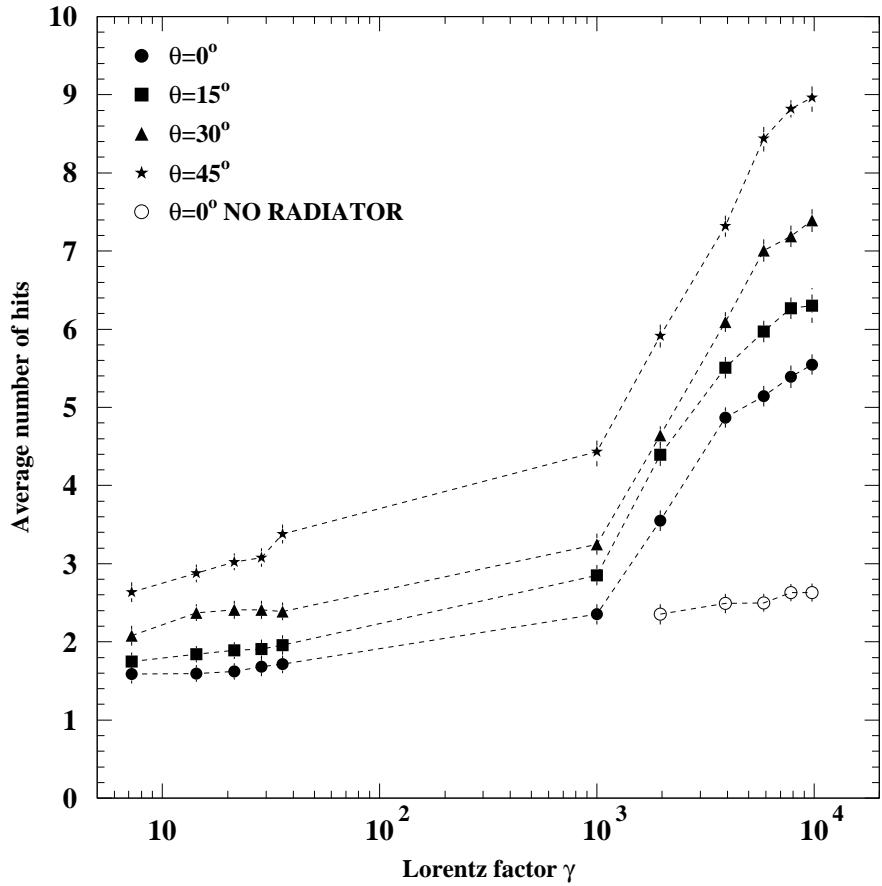}}
\end{center}
\caption{Average number of hits as a function of the Lorentz factor
$\gamma$ for several beam crossing angles. The dashed lines are drawn 
to guide the eye.}
\label{fig:trdcalib}
\end{figure}

A reduced scale prototype of the TRD was exposed
to a pion/electron beam~\cite{trd2} to evaluate the 
detector response function.
The physical observable related to the energy of the 
particles crossing the detector is the total number 
of hits produced in the proportional tubes.
For a sample of particles crossing the detector with 
a fixed energy and at a fixed angle, the number of hits 
are roughly Poisson distributed, with an average value 
of few units, which depends on the beam energy and
crossing angle.  
In Fig.~\ref{fig:trdcalib} the average number of measured
hits in the proportional tubes is plotted as a function of the 
Lorentz factor $\gamma$ of the incident particles, 
for several beam crossing angles.
Below the threshold value ($\gamma = 10^{3}$) there is only 
the contribution of the ionization energy loss; 
for $10^{3} < \gamma < 10^{4}$, the TR contribution is also present 
and the average number of hits increases logarithmically 
with $\gamma$.

  
\section{Data selection and analysis}
\label{sec:data}

\begin{figure}[t]
\begin{center}
\resizebox{13.0cm}{7.0cm}{\includegraphics{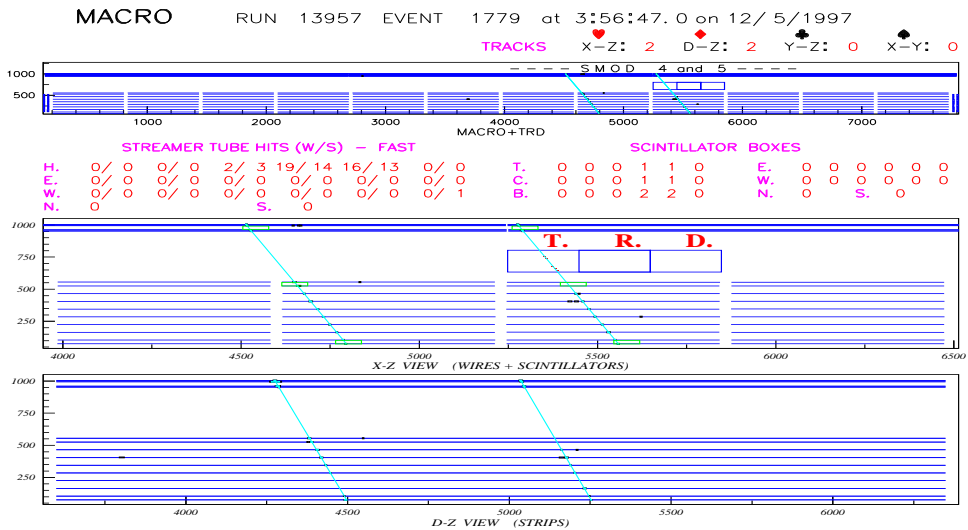}}
\resizebox{13.0cm}{7.0cm}{\includegraphics{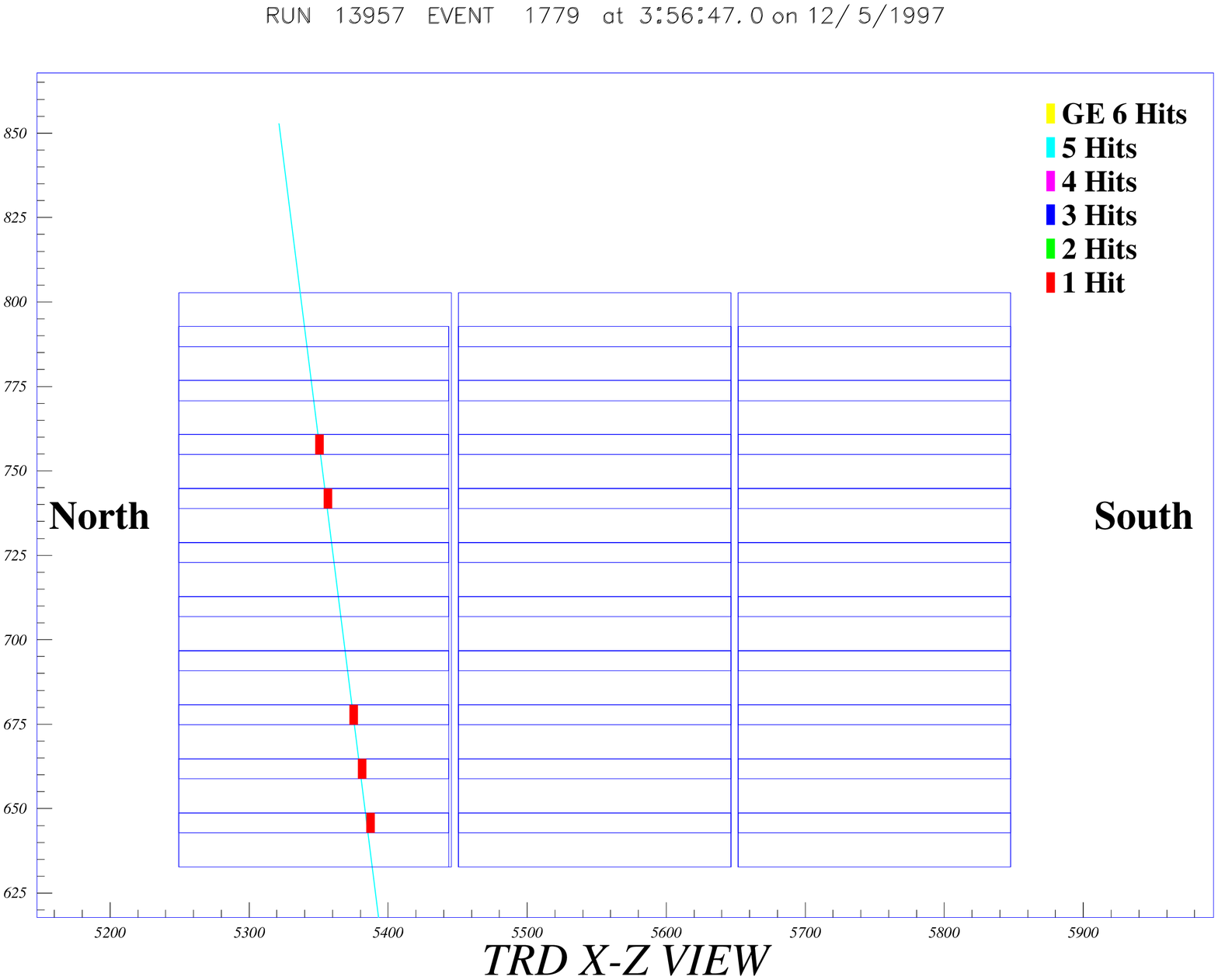}}
\end{center}
\caption{Display of a double muon event.} 
\label{fig:evdisp} 
\end{figure}

We considered the data collected by all the
three MACRO TRD modules during the acquisition period
from April 1995 to December 2000. 
Two classes of events were analyzed:
\begin{enumerate}
\item{``single muons'', i.e. events with one muon in MACRO crossing
    one of the TRD modules;}
\item{``double muons'', i.e. events with two muons in MACRO and only
    one muon crossing one of the TRD modules, like the one shown in
    figure~\ref{fig:evdisp}.}
\end{enumerate} 

To evaluate the muon energies, we associated to each muon track 
the hits produced in the TRD proportional tubes. 
The muon was tracked with the standard MACRO procedure~\cite{macro},
which uses the information of the streamer tube system. 
The distribution of the distances of the TRD hits 
from the expected position, calculated with the muon reconstructed
track, has a gaussian shape with a standard deviation 
$\sigma \approx 2 \units{cm}$. 
We associated to each muon all the proportional tubes hits
within $3 \sigma$ from the track.
To avoid badly reconstructed tracks, 
only muons crossing at least three
layers of streamer tubes in the lower part of MACRO were used. 
Muons accompanied with electromagnetic showers initiated in the 
rock surrounding the detector were also discarded.
Since the TRD was calibrated (see Fig.~\ref{fig:trdcalib}) 
with particles crossing all the ten layers of 
proportional tubes at angles below $45 \degrees$,
in the present analysis only muons  
crossing the whole detector with zenith angle 
smaller than $45 \degrees$ were included.  

Runs in which the TRD modules were affected by stability problems 
or malfunctioning were discarded. Also runs with 
reconstructed muon rates differing more than three standard 
deviations from the average values were disregarded, 
as well as those runs whose duration was less than $1 \units{h}$.
The final data samples consist of $250290$ single muons
and $17942$ double muons, for a total life time of $2586$ days 
(see Table~\ref{tab:summary}).

\begin{figure}
\begin{center}
\resizebox{13.0cm}{13.0cm}{\includegraphics{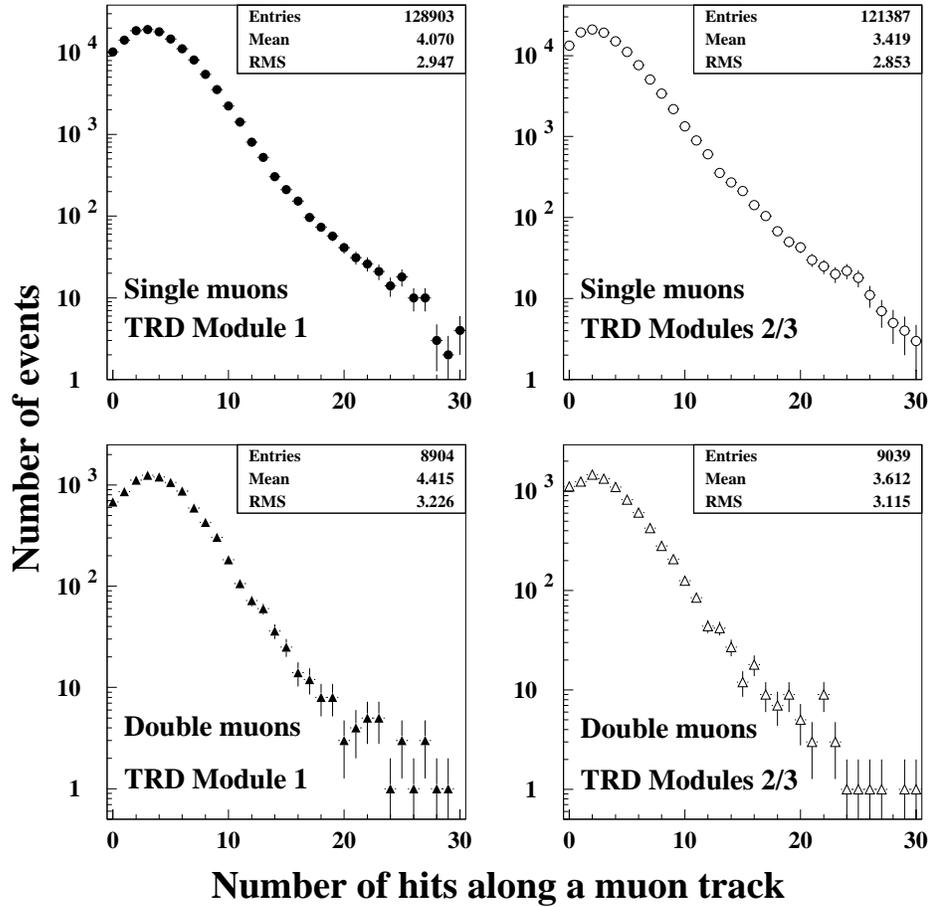}}
\end{center}
\caption{Hit distributions $N(k)$ vs. $k$ 
for the final single and double muon data samples. 
Since the read-out electronics of the first TRD module 
was different from that of the second and third modules, 
the data data samples from the first module and the ones 
from the second and third modules were analyzed separately.
}
\label{fig:hitdist}
\end{figure}

Fig.~\ref{fig:hitdist} shows the distributions of the
number of hits produced in the TRD proportional tubes along the 
muon tracks for the final data samples. Since the second 
and third TRD modules were equipped with a different read-out 
electronics from that of the first module, two different 
response functions (described in the next section) were necessary 
to analyze the data samples from the first module and 
those from the second and third modules. The results of the
two analysis were then combined.


\section{The TRD response function}
\label{sec:trdresp}


\subsection{Evaluation of the TRD response function}

We define $N(k \mid h, \theta)$ as the distributions of the TRD hits 
for a sample of muons with zenith angle $\theta$ that crossed a 
rock thickness $h$. The rock thickness $h$ was calculated
from the direction $(\theta, \phi)$ using the Gran Sasso 
map~\cite{gsmap} and was converted into standard rock according 
the prescriptions of~\cite{stroc}. 
These distributions can be related to the residual 
muon energy spectra $N(E \mid h, \theta)$ by:
\begin{equation}
N(k \mid h, \theta) = \sum_{E}~p(k \mid E, \theta) N(E \mid h, \theta)
\label{eq:hit}
\end{equation}
where $p(k \mid E, \theta)$ is the TR detector response function,
i.e. the probability to observe $k$ hits along the track of a 
muon with underground energy $E=E_{\mu}$, crossing the detector 
at a zenith angle $\theta$.

To reconstruct the muon energy spectrum $N(E \mid h, \theta)$ 
starting from the measured hit distributions (Fig.~\ref{fig:hitdist}),
once the TRD response function  $p(k \mid E, \theta)$ is known,
eq.~\ref{eq:hit} must be inverted.
We derived two matrices $p(k \mid E, \theta)$ 
(one for the first TR module and another for the second and third
modules) on the basis of the calibration data taken by a 
reduced scale prototype exposed to a pion-electron 
test beam at CERN \cite{trd2}.

We simulated, with a full GEANT-based~\cite{brun} Monte Carlo, 
a muon sample distributed according a flat energy and solid angle 
spectrum (i.e. $\frac{d^{2}N}{d E~d \Omega} = constant$). 
On the basis of the calibration data shown in 
Fig.~\ref{fig:trdcalib}, seven energy bins 
and four angular bins were defined. The first 
energy bin covers the range from $0$ to $50 \units{GeV}$ 
(i.e. $0< \gamma < 500$), and the last one covers the range 
above the TRD saturation 
($E \geq 1 \units{TeV}$, i.e. $\gamma \geq 10^{4}$). 

The number of hits produced in the TRD by a simulated muon 
of energy $E$ crossing the detector at zenith angle $\theta$ 
was extracted by the corresponding calibration data set.
The TRD response function (a $31 \times 7 \times 4$ matrix) 
was calculated as:
\begin{equation}
p(k \mid E, \theta) = 
\frac{\mathcal{N}(k \mid E, \theta)}{\mathcal{N}(E, \theta)}
\end{equation}
where $\mathcal{N}(k \mid E, \theta)$ is the number of 
simulated muons with residual energy $E$ and zenith angle 
$\theta$, producing $k$ hits, with $0 \le k \le 30$ and 
$\mathcal{N}(E, \theta)$ is the total number of simulated muons 
with energy $E$ crossing the TRD at a zenith angle $\theta$.
The simulated data had the same format 
of experimental data and were processed by the same analysis 
tools used for real data.


\subsection{Check of the TRD response function}

The accuracy and the time stability of the TRD response function 
also needed to be taken into account. 
Although the nominal gas gain of the TRD 
proportional tubes was the same of the tubes used in the prototype, 
during the long time-scale data acquisition period their operating 
conditions were affected by fluctuations. This was 
due to drifts of some 
important parameters, like the gas mixture composition, the
atmospheric pressure and the temperature.     

A first check of the TRD response function for energies below 
the threshold for the emission of TR was made using stopping muons, 
i.e. muons that are absorbed in the lower part of MACRO, 
that can be easily tagged by imposing simple geometrical cuts. 
Since the average residual energy of these muons is below 
a few $\units{GeV}$, we simulated a sample of muons with 
a flat energy spectrum up to $10 \units{GeV}$. 
The same algorithm for the selection of stopping muons 
was applied to both the real and the simulated data samples.
The measured hit distributions of stopping muons are 
well reproduced by the Monte Carlo simulation
(see Table~\ref{tab:mustop}), 
thus confirming the reliability of the TRD calibration 
in the low energy region.  

\begin{figure}
\begin{center}
\resizebox{13.0cm}{13.0cm}{\includegraphics{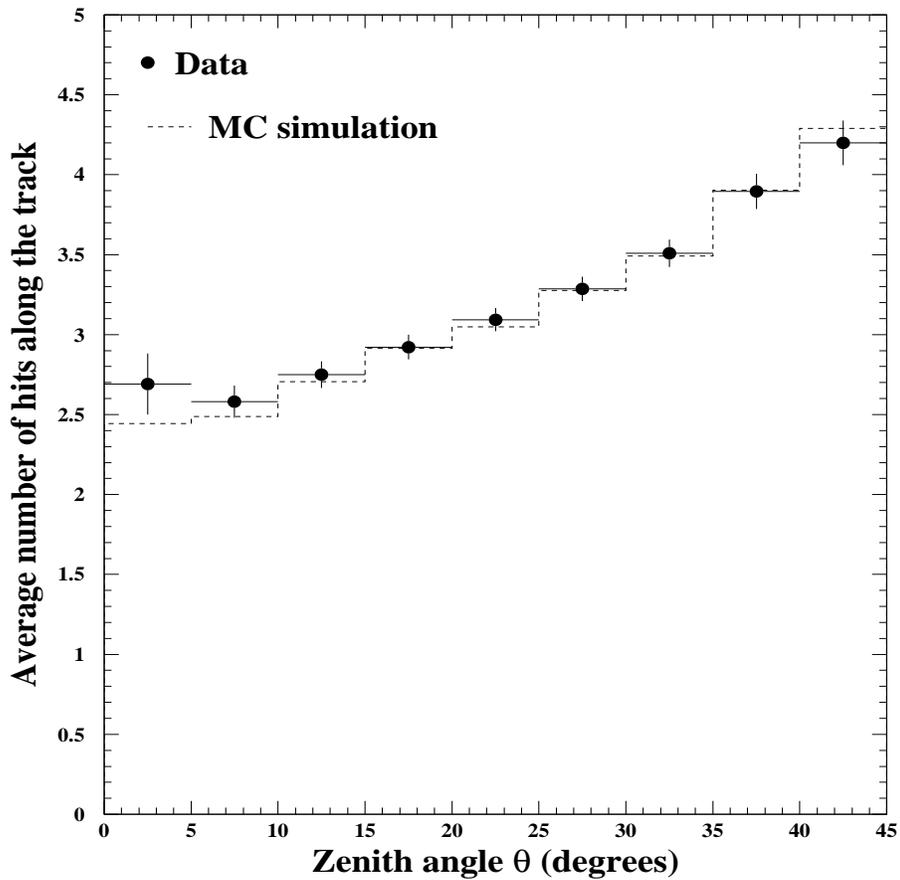}}
\end{center}
\caption{Check of the TRD response function by means of muons
crossing the region of the third module equipped with aluminum foils.
Data ($\bullet$) are compared with the Monte Carlo 
predictions(dashed line). }
\label{fig:trdal}
\end{figure}

A further check of the detector response function was 
performed using the data from the part of the third 
module equipped with aluminum foils. 
Since the aluminum foils absorbed the TR produced 
in the upstream radiator, this data sample allows to check 
the detector response for muons releasing energy in the
proportional tubes only by ionization. 
To take into account the weak dependence of the TRD response 
on ionization (see Fig.~\ref{fig:trdcalib}), 
we simulated a sample of muons with the characteristic 
energy spectrum at MACRO depth (eq.~\ref{eq:undspec}) 
and with the same angular distribution as real data.
In Fig.~\ref{fig:trdal} the values of the average 
number of hits along the muon track versus the muon zenith 
angle from the Monte Carlo and from the real data are shown.  
The $\chi^2/D.o.F.$ value (evaluated using only 
statistical errors) is $3.3/9$, showing that there is 
a good agreement between the data and the simulation.  

From this analysis we estimated the systematic 
uncertainty arising from the fluctuations of the detector 
operating conditions. For each of the four angular bins 
used for evaluating the TRD response function, we
compared the measured hit distributions 
$N(k \mid \theta)$ with the simulated ones.
We remarked that the differences between the 
experimental average values and the Monte Carlo
predictions are within $\pm 5\%$.
For this reason, to take into account the fluctuations
of the detector operating conditions, we associated to 
the average values of each hit distribution of the calibration
data set, an uncertainty of $\pm 5\%$. 
These uncertainties will propagate to the systematic 
errors affecting the measurement of the residual 
energy of muons. 

The systematic errors quoted in the present analysis 
(see sec.~\ref{sec:spectra})  are greater than the ones quoted in 
our previous work~\cite{trd3}. When that analysis
was performed, the TRD module equipped with aluminum foils
was still not taking data. For this reason, we could estimate the
systematic uncertainties only on the basis of the beam test
data and we decided to associate to the calibration data 
set an uncertainty of $\pm 2\%$, that is
lower than the value we adopted for this analysis.


\section{Evaluation of the muon energy spectra}
\label{sec:spectra}

As shown in section~\ref{sec:trdresp}, the residual energy spectra 
of underground muons are related to the measured hit distributions 
in the TRD by eq.~\ref{eq:hit}. 

Two different methods were applied to reconstruct the underground 
muon energy spectra by inversion of eq.~\ref{eq:hit}. 
The first approach uses the same unfolding procedure described 
in~\cite{trd3}. In the second approach a parameterization for 
the muon energy spectrum was assumed and the parameters 
were derived using a best fit procedure.


\subsection{The unfolding procedure}
\label{sec:unfo}

Unfolding techniques are widely applied in problems where 
matrix inversion is required~\cite{dago}. To reconstruct 
the energy spectra of single and double muons starting from 
the hit distributions of Fig.~\ref{fig:hitdist}, we applied 
an iterative unfolding procedure~\cite{dago,mazz}.  
As an initial energy spectrum (used as a starting point for 
the unfolding procedure), we assumed eq.~\ref{eq:undspec} 
for both the single and the double muon events, with parameters: 
$\alpha=3.7$, $\beta=0.383~10^{-3} \units{hg^{-1}cm^{2}}$ 
and $\epsilon=620 \units{GeV}$.
The final reconstructed spectrum does not depend on the choice 
of the initial spectrum. It only affects the time needed for 
the procedure to converge~\cite{dago,mazz}.

The energy spectrum reconstructed after each iteration is used 
as input for the next iteration. At the end of each iteration 
a $\chi^{2}$ test is performed between the reconstructed energy 
spectra and the input energy spectra. The $\chi^{2}$ is defined as:
\begin{equation}
\chi^{2} = \sum_{i,j,k} 
\cfrac{\left[ N_{n}(E_{i} \mid h_{j}, \theta_{k}) - 
N_{n-1}(E_{i} \mid h_{j}, \theta_{k}) \right]^{2}}
{\sigma_{n}^{2}(E_{i} \mid h_{j}, \theta_{k}) +
\sigma_{n-1}^{2}(E_{i} \mid h_{j}, \theta_{k})} 
\end{equation}
where $N_{n-1}(E_{i} \mid h_{j}, \theta_{k})$ and 
$N_{n}(E_{i} \mid h_{j}, \theta_{k})$ are the energy spectra 
reconstructed after the $(n-1)$-th and the $n$-th iteration
and $\sigma_{n-1}(E_{i} \mid h_{j}, \theta_{k})$, while
$\sigma_{n}(E_{i} \mid h_{j}, \theta_{k})$ are the associated 
errors.
The iterative procedure stops when the energy distribution 
reconstructed after the $n^{th}$ iteration has a $\chi^{2}$ 
probability greater than $99\%$ to be equivalent to the one 
reconstructed after the $(n-1)^{th}$ iteration.  
The unfolding procedure was separately applied to the data 
samples from the first and from the second and third TRD modules 
and the results were combined.

\begin{figure}
\begin{center}
\resizebox{13.0cm}{13.0cm}{\includegraphics{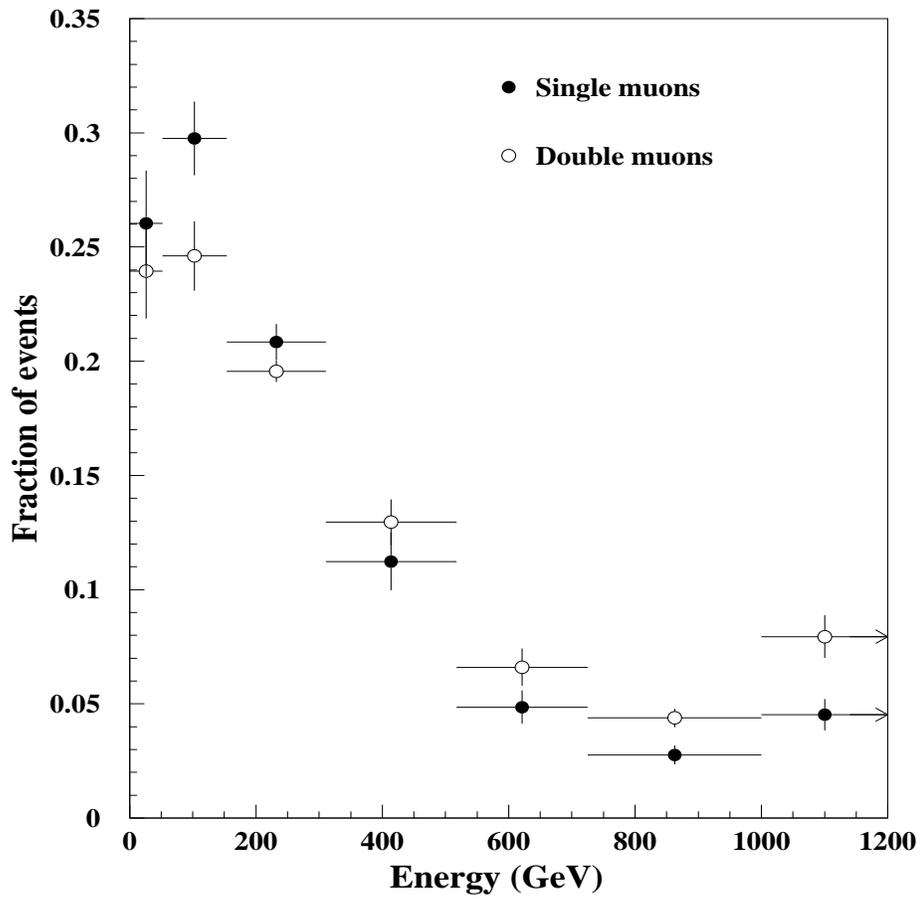}}
\end{center}
\caption{Reconstructed differential energy spectra of single and
double muons. Statistical and systematic errors have been added
in quadrature.}
\label{fig:diffspectra}
\end{figure}

Fig.~\ref{fig:diffspectra} shows the reconstructed differential 
energy spectra of single and double muons at the Gran Sasso 
underground laboratory depth. The error bars in the figure 
were calculated by adding in quadrature statistical and
systematic errors.
The systematic errors are originated by fluctuations of the 
TRD response function, and were estimated by a $+5\%$ and a $-5\%$ 
variation of the calibration data. The unfolding procedure 
was replicated using the ``perturbed'' detector response 
functions; the systematic uncertainty was evaluated as the difference 
between the two results.
 
The unfolding procedure reconstructs the shapes of 
the spectra up to $1 \units{TeV}$. For energies greater than 
$1 \units{TeV}$, where the TRD response is saturated, 
the spectral shapes cannot be reconstructed, and
only the number of events can be evaluated. The average value 
of the energy of underground muons with energy above threshold, 
$\langle E \rangle_{E>E_{0}}$, was calculated from 
eq.~\ref{eq:undspec} as:
\begin{equation}
\label{eq:enocut}
\langle E \rangle _{E>E_{0}} = \frac{\alpha - 1}{\alpha - 2} \cdot
\left( E_{0} + \frac{\epsilon (1-e^{-\beta h})}{\alpha - 1} \right).
\end{equation}

The average muon energy was evaluated as:
\begin{equation}
\langle E \rangle = 
(1-f) \langle E \rangle_{E< E_{0}} + f \langle E \rangle_{E> E_{0}}
\end{equation} 
where
$\langle E \rangle_{E<E_{0}}$ is the average energy 
of muons with energies up to $E_{0}=1 \units{TeV}$, 
$f$ and $\langle E \rangle_{E>E_{0}}$ are the fraction 
and the average energy of muons with energies greater 
than $1 \units{TeV}$, respectively.  
For a $3\%$ variation of the parameters 
$\alpha$, $\beta$ and $\epsilon$, as is typically 
quoted by various authors (e.g.~\cite{batt}), 
the uncertainties on
the average muon energies are less than $1\%$ 
and are significantly smaller than our quoted errors. 
The results are shown 
in Table \ref{tab:energy}. The average energy of single
muons and double muons are respectively of 
$270 \pm 3 (stat.) \pm 18 (syst.) \units{GeV}$ and
$381 \pm 13 (stat.) \pm 21 (syst.) \units{GeV}$.
The single muon result is not in contradiction with
the result of our previous analysis~\cite{trd3}, where
the systematic effects were underestimated.


\subsection{Best fit of the spectral indices}
\label{sec:fit}

An alternative way for evaluating the muon energy spectra from 
the measured hit distributions is that of assuming
eq.~\ref{eq:undspec} as an analytic description for the spectra, 
and to derive the parameters using a best-fit. 
Substituting the trial energy spectrum into eq.~\ref{eq:hit} 
and summing over the zenith angles we get:
\begin{equation}
N (k \mid h) = \sum_{E} \sum_{\theta} p(k \mid E, \theta)~
c(h, \theta) \left[ E + \epsilon (1 - e^{-\beta h}) \right]^{-\alpha}
\label{eq:hit2}
\end{equation}
where the normalization constants $c(h, \theta)$ represent the
number of muons detected in each bin of depth and zenith angle.
We fixed the values of the $\beta$ and $\epsilon$ parameters to
$\beta = 0.383~10^{-3} \units{hg^{-1}cm^{2}}$ and 
$\epsilon = 620 \units{GeV}$~\cite{lipari}; $\alpha$ was left 
as the free parameter for the single and double muon data.  

For each value of $\alpha$ in the range from $2$ to $6$ and step $0.1$ 
we built a set of hit distributions $\hat{N}(k \mid h)$ 
according eq.~\ref{eq:hit2}, and for each set of distributions 
we evaluated the $\chi^{2}$ as:
\begin{equation}
\chi^{2} = \sum_{TRD~modules} \sum_{h} \sum_{k} 
\frac{[ N(k \mid h) - \hat{N}(k \mid h)]^{2}}
{\sigma^{2}(k \mid h) + \hat{\sigma}^{2}(k \mid h)}
\end{equation}
where $N(k \mid h)$ is the measured hit distribution 
for muons (single or double) crossing a rock slant depth $h$; 
$\sigma(k \mid h)$ and $\hat{\sigma}(k \mid h)$ are the 
Poissonian errors on $N(k \mid h)$ and $\hat{N}(k \mid h)$, 
respectively. 

The curves representing the $\chi^{2}$ as a function of 
$\alpha$ are continuous both for single and double muons and show 
a well defined minimum. The value of the best fit spectral 
index for single muons is 
$\alpha_{1} = 3.79 \pm 0.02 (stat) \pm 0.11 (syst)$, 
with $\chi^{2}_{min} / d.o.f. = 1.51$,
while for the double muons is 
$\alpha_{2} = 3.25 \pm 0.06 (stat) \pm 0.07 (syst)$, 
with $\chi^{2}_{min} / d.o.f. = 0.55$.
The statistical error on $\alpha$ was evaluated using the 
values $\alpha_{L}$ and $\alpha_{R}$ corresponding to 
$\chi^{2} = \chi^{2}_{min} + 1$. 
The systematic error was estimated from the positions of 
the new minima of $\alpha$ obtained using the two TRD 
response functions evaluated from the sets of 
``perturbed'' calibration data.

The result obtained for the single muon spectral index is 
consistent with that obtained from the MACRO measurement of 
the underground muon intensity as a function of the rock 
depth~\cite{ih}. It is also consistent with the results 
of the NUSEX experiment~\cite{nusex}, that found a 
spectral index of $\alpha = 3.91^{+0.50}_{-0.36}$ for a sample 
of events mainly composed by single muons.


\subsection{Comparison between the results}   

\begin{figure}
\begin{center}
\resizebox{13.0cm}{15.0cm}{\includegraphics{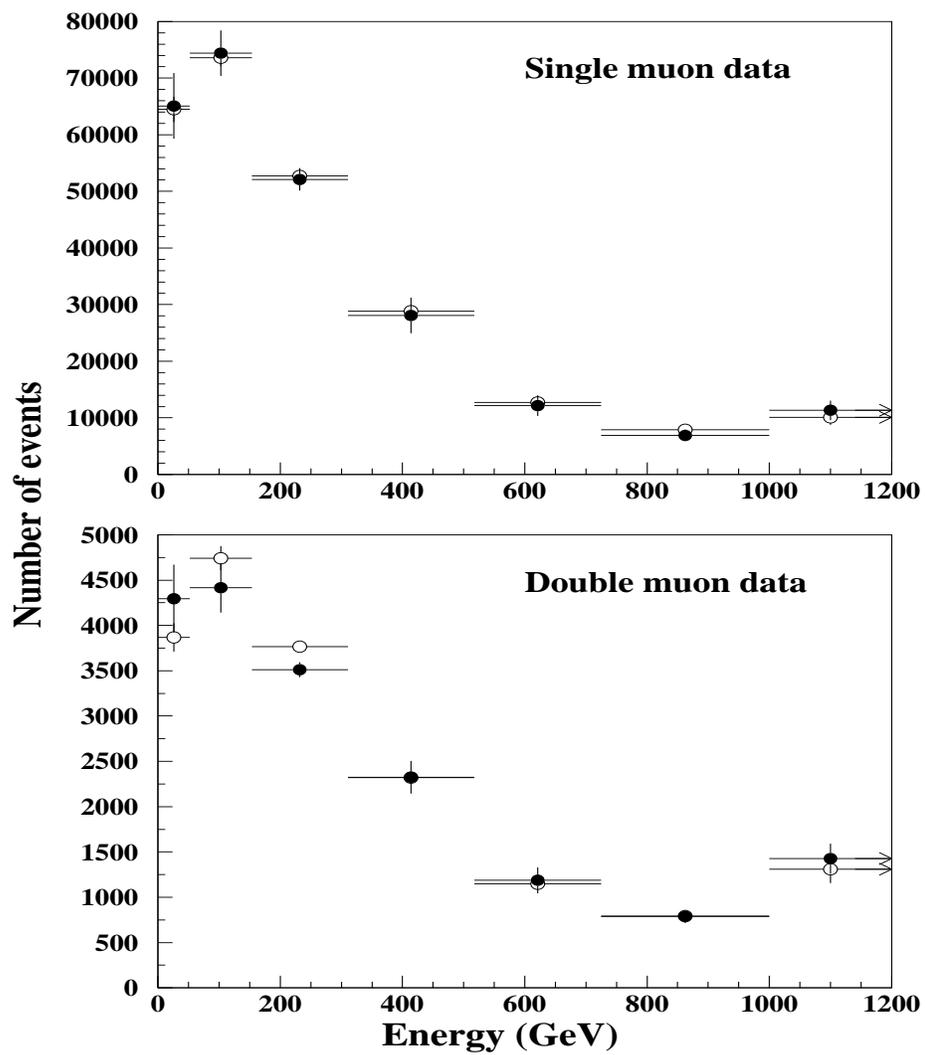}}
\end{center}
\caption{Reconstructed differential energy spectra of single and
double muons. Results from the unfolding procedure (black dots) 
are compared with those from the best fit procedure (open circles).
Statistical and systematic errors have been added
in quadrature.}
\label{fig:confronto}
\end{figure}

Fig.~\ref{fig:confronto} shows a comparison between 
the differential muon energy spectra reconstructed 
with the two methods described in sections~\ref{sec:unfo} 
and~\ref{sec:fit} both for single and double muons.  
In the case of single muons we find a value of 
$\chi^{2} / d.o.f. = 1.3/7$, 
while in the case of double muons we find 
$\chi^{2} / d.o.f. = 8.5/7$. 
In the case of single muon data, there is a good 
agreement between the result obtained using the unfolding 
procedure and that obtained using the best fit method. 
In the case of double muon data, there are some 
discrepancies, especially for low residual energies.

To understand these difference, it should be noted 
that the result of the unfolding procedure does not depend 
on the trial spectrum. The spectrum 
used for the fit (eq.~\ref{eq:undspec}) is derived assuming 
that the surface muon energy spectra obey a power 
law with spectral index $\alpha$, and in the muon 
propagation formula, $\epsilon$ and $\beta$ are constant, 
i.e. the cross sections for the radiative processes do 
not depend on the muon energy.

If the correct expression (eq.~\ref{eq:musurf}) is 
assumed for the surface energy spectrum and the dependence 
of $\epsilon$ and $\beta$ on the muon energy is taken 
into account, it is impossible to obtain
an analytical expression 
for the underground muon spectrum. 
Hence, eq.~\ref{eq:undspec} represents
only a useful parameterization for the muon 
underground spectra, whose accuracy is limited by 
the hypotheses from which it has been derived. 
The difference between our results obtained using 
the two techniques mentioned in sections~\ref{sec:unfo}
and~\ref{sec:fit} are ascribed to 
these approximations. We can conclude that the 
parameterization is not 
completely suitable for describing the underground 
muon energy spectra, especially for events with 
large underground multiplicity.


\section{Monte Carlo simulation}
\label{sec:MC}

As shown in section~\ref{sec:undmuon}, the underground 
muon energy spectrum is related to the primary cosmic ray 
spectrum. In order to investigate the relationship between 
the underground muon energy and the primary cosmic ray 
spectrum and composition, we performed a Monte Carlo 
simulation using different composition models for primary 
cosmic rays.

The interactions of cosmic rays in the atmosphere were
simulated with the HEMAS code~\cite{hemas}, while the 
process of muon propagation in the rock was simulated 
using the PROPMU code~\cite{lipari}. Two extreme 
composition models were assumed for primary cosmic rays:
\begin{itemize}
\item[a.]{the ``light model''~\cite{light}, i.e. a proton-rich
composition model;} 
\item[b.]{the ``heavy model''~\cite{heavy}, i.e. a Fe-rich 
composition model.}
\end{itemize}
These models assume that cosmic rays are composed by 
five main mass groups ($p$,$He$,$CNO$,$Mg$ and $Fe$). 
The energy spectrum of each component is described 
by means of power laws given by:
\begin{equation}
\label{eq:compo}
 \frac{dN_i}{dE} = \left\{
 \begin{array}{ll}
 K_{1}E^{-(\gamma_{1}+1)} & E \leq E_{cut} \\ \\
 K_{2}E^{-(\gamma_{2}+1)} & E > E_{cut} \\
 \end{array}
 \right. 
\end{equation}
In eq.~\ref{eq:compo} there are 5 parameters for each 
primary component (i.e. the normalization constants 
$K_1$ and $K_2$, the spectral indices $\gamma_1$ and $\gamma_2$ 
and the cutoff energy $E_{cut}$).
These parameters are not independent: usually the constant 
$K_2$ is expressed as a function of the others by imposing 
the continuity of the function $dN_{i}/dE$ at $E=E_{cut}$. 
The all-particle spectrum is evaluated by adding the
contributions from all the mass groups.
In Table \ref{tab:models} the values of all the parameters are 
summarized for each component of the primary cosmic rays in 
both the composition models we adopted. 
The light model 
is different from the heavy model because there is an 
extra-component of protons. 

\begin{figure}
\begin{center}
\resizebox{12.0cm}{12.0cm}{\includegraphics{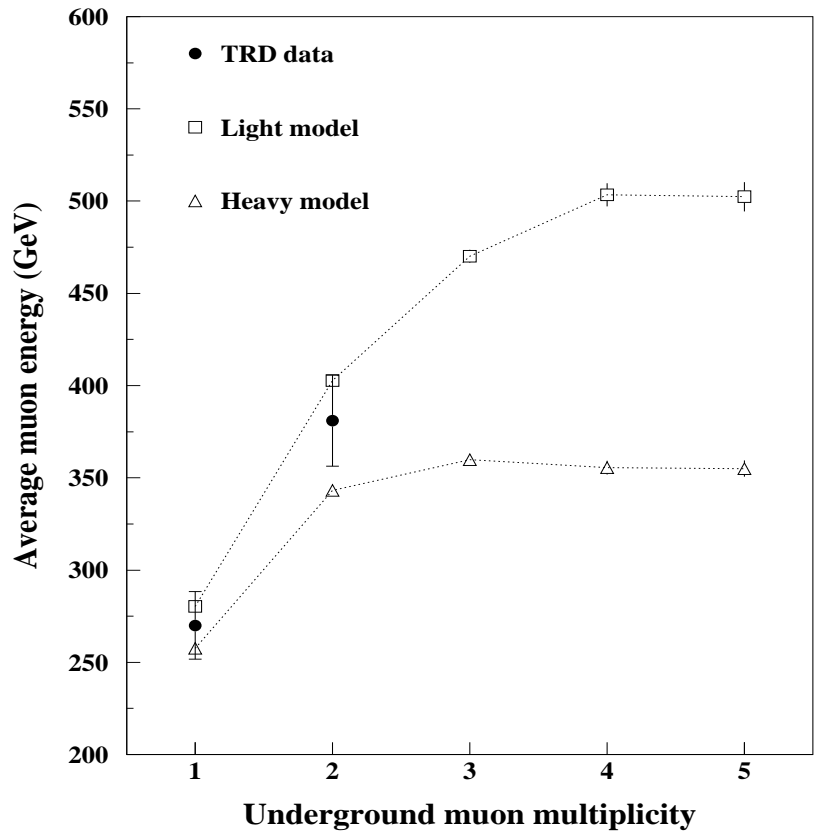}}
\end{center}
\caption{Average energies of underground muons as a fuction of 
their multiplicity according the predictions obtained assuming 
the light and heavy composition models. The Monte Carlo predictions
are compared with the results from the unfolding procedure shown 
in Table~\ref{tab:energy}. The error bars represent 
the sum in quadrature of systematic and statistical errors.}
\label{fig:evsnmu}
\end{figure}

Fig.~\ref{fig:evsnmu} shows the predictions obtained 
using the light and heavy composition models for the 
average energy of the underground muons as a function 
of the muon multiplicity. The gap between the
predictions from the two extreme models increases 
with the muon multiplicity. The two dark points represent 
our measured values, averaged on the whole rock 
depth range from $3000$ to $6500 \units{hg/cm^{2}}$. 

\begin{figure}
\begin{center}
\resizebox{12.0cm}{16.0cm}{\includegraphics{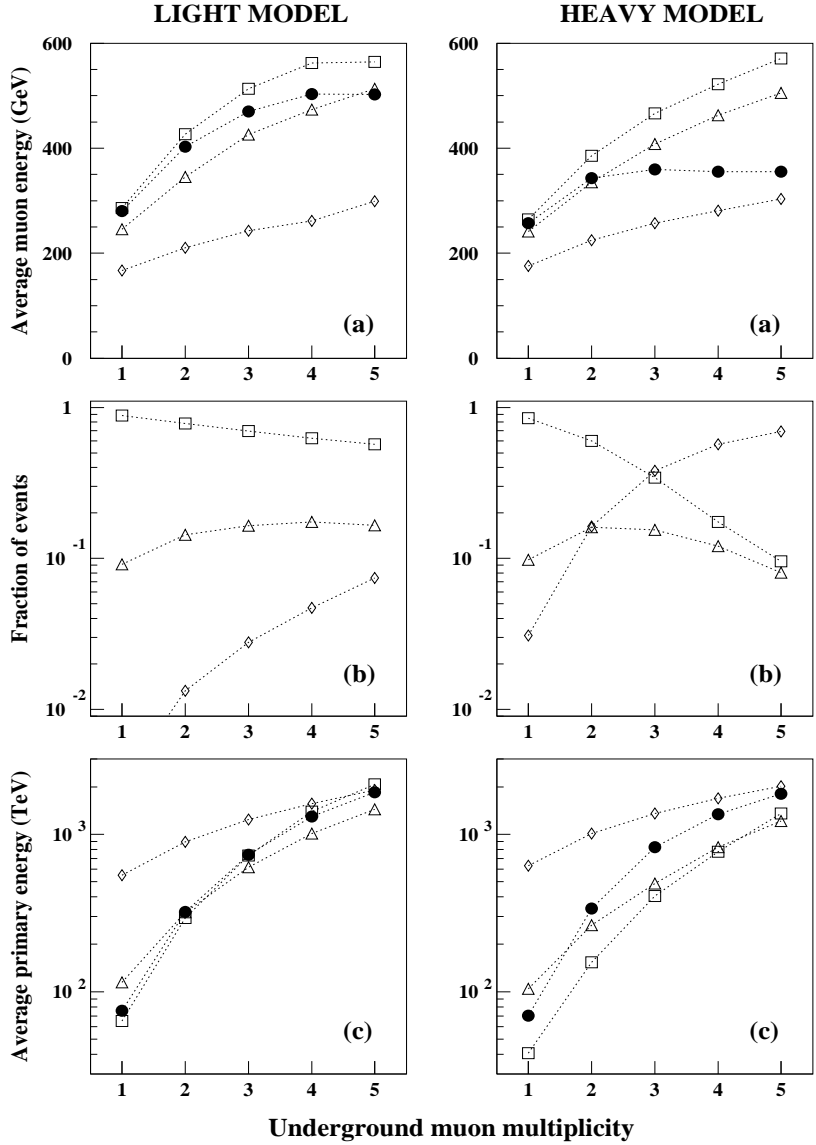}}
\end{center}
\caption{Physical quantities related to the underground muons
produced by the main components of the primary cosmic
rays ($\Box = p$, $\vartriangle = He$, $\lozenge = Fe$, $\bullet$ = all
particles) plotted as a function of the muon multiplicity:
a) average energies of underground muons; 
b) fraction of underground muons; 
c) average energies of the parent cosmic rays.
Light (left panels) and heavy (right panels) composition models
were separately considered.}
\label{fig:mcsummary}
\end{figure}

To explain the average energy behavior, in Fig.~\ref{fig:mcsummary} 
we plotted, for the two composition models and as a 
function of the underground muon multiplicity, for
the $p$, $He$ and $Fe$ components: 
$a)$ the average energies of underground muons; 
$b)$ the relative contribution of these three mass 
groups and $c)$ the average energies of the parent cosmic rays 
to events for a fixed underground muon multiplicity.
In $a)$ and $c)$ the values for the all-particle spectrum
are also shown. 
From this figure the following conclusions can be drawn:
\begin{itemize}
\item[a.] single muons are originated mainly by primary protons;
\item[b.] the contributions of primary heavy elements become 
relevant when the muon multiplicities increase, while the 
contribution of protons decreases;
\item[c.] in the "heavy" composition model the contribution of
$Fe$ nuclei is relevant from low multiplicities (it is $\sim 15\%$ for
$N_{\mu}=2$ and tends to $100\%$ for high multiplicities);
\item[d.] in the light model, the main contribution is 
always that of protons;
\item[e.] for a fixed muon multiplicity, muons originated 
by light primaries are more energetic than muons originated 
by heavy primaries. 
\end{itemize}

From the simulation we get that
the average energies of underground muons produced 
by each component are almost independent on 
the primary composition model. 
This is due to the fact that the differences between the spectral 
indices of the primary components in the models we are examining 
are small for energies below the ``knee'', as can be seen from 
table \ref{tab:models}. As a consequence, the muons produced 
e.g. by He nuclei have the same average energy, independently 
on the choice of the composition model. 
Hence, the differences between the predictions of the cosmic ray 
composition models are mainly due to differences in the weights 
of the various components in each model and not to differences 
in their spectra. 

From Fig.~\ref{fig:mcsummary}$c$, the primary energy needed 
to produce one underground muon at the MACRO depth
is smaller than $100\units{TeV}$, 
while the primary energy needed to produce two underground muons 
is about $300-400\units{TeV}$. 
This means that an analysis of events with only one or two 
underground muons is sensitive to the energy region below 
the knee of the primary spectrum. 

\begin{figure}
\begin{center}
\resizebox{12.0cm}{8.0cm}{\includegraphics{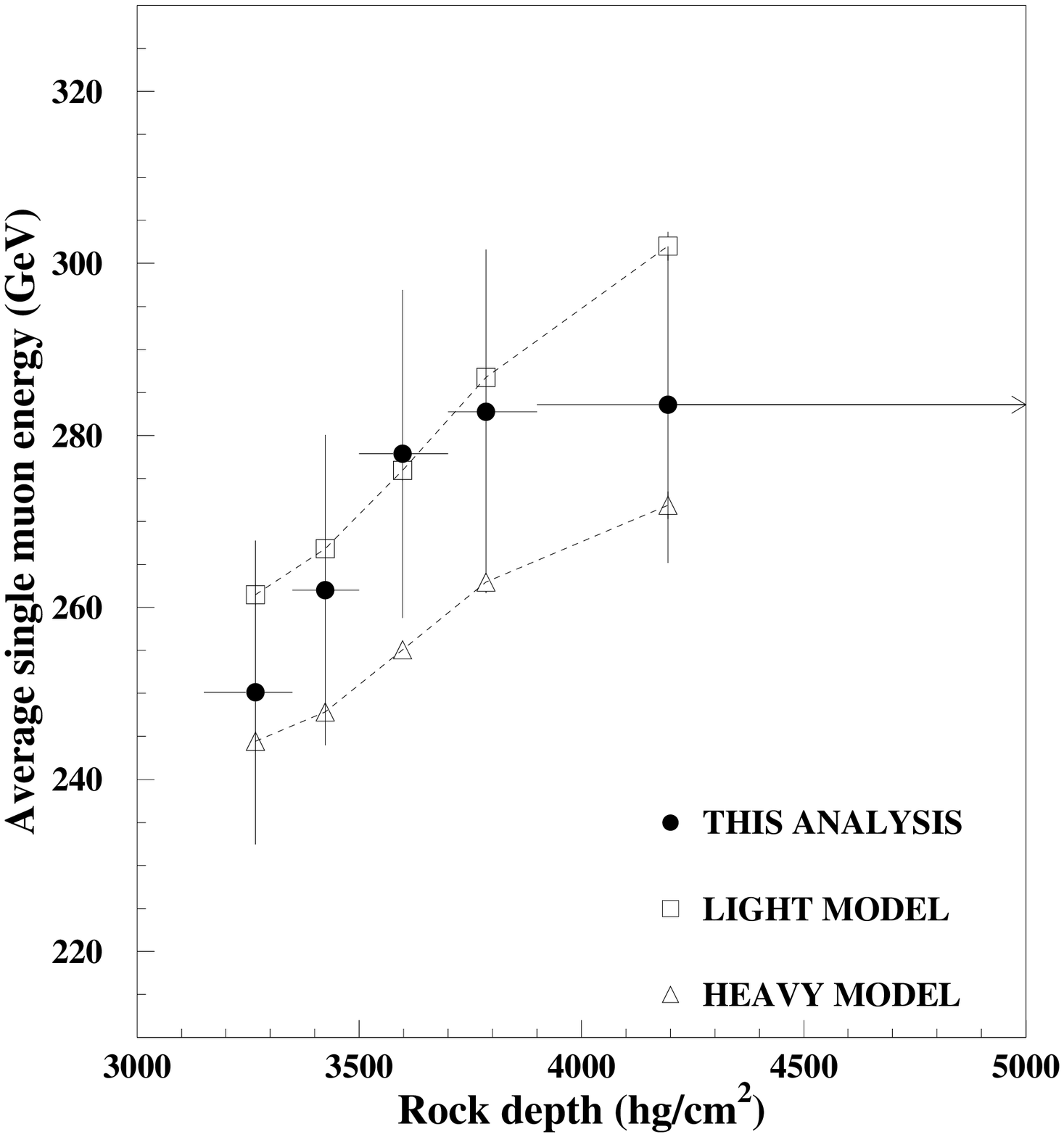}}
\resizebox{12.0cm}{8.0cm}{\includegraphics{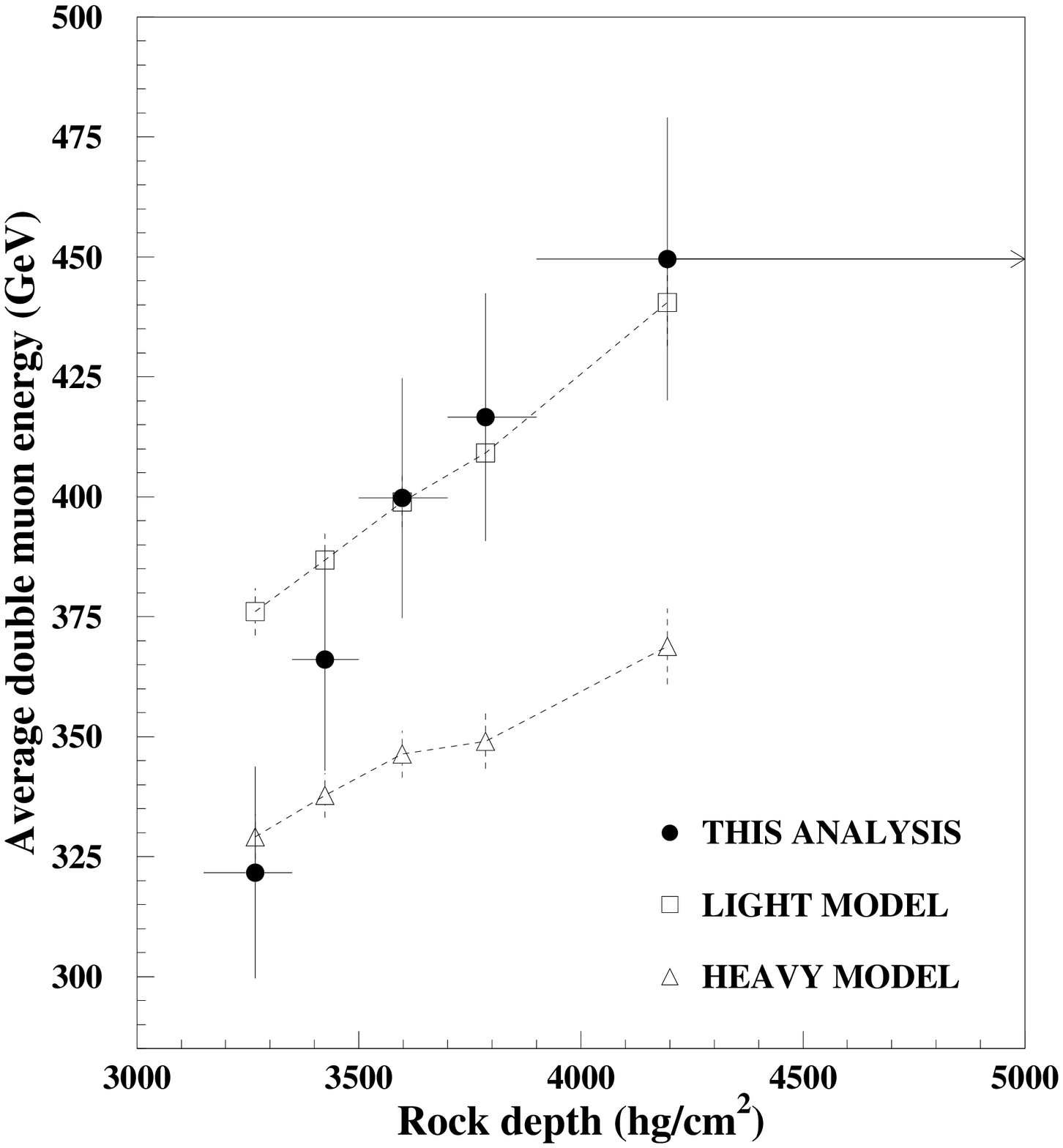}}
\end{center}
\caption{Average energies of single and 
double muons as a function
of the rock depth. 
Statistical and systematic errors have been added in quadrature.
The horizontal bars represent the width of the $h$ bins, while the
central value of each bin corresponds to the average value of $h$
for that bin. The last bin extends up to $6500 \units{hg/cm^{2}}$.
Results from the unfolding procedure 
are compared with the Monte Carlo predictions.}
\label{fig:datamc}
\end{figure}

Our measurements were compared with the predictions
from the simulation.
Fig.~\ref{fig:evsnmu} shows a comparison between the results 
obtained applying the the analysis technique described in 
sec.~\ref{sec:unfo} and the Monte Carlo predictions concerning 
the average energies of single and double muons, while
in Fig.~\ref{fig:datamc} the average energies of single and
double muons are plotted as a function of the rock thickness 
crossed by muons. 


\subsection{Discussion of the results}
\label{sec:discussion}

Single muon data are not in contradiction
with the predictions from both the composition models.
In fact, since the energy gaps between the predictions
from the two extreme composition models are of the same order
of magnitude as the error bars, single muons
do not allow one to discriminate between 
different cosmic ray composition models.   
The sensitivity of this measurement to the primary cosmic
ray composition is strongly limited by systematic errors 
associated to the TRD response function.   

Also in the case of double muons, our data do not allow
to perform a discrimination between the cosmic ray
composition models. In this case, although the errors
associated to the average muon energies are smaller than
the gap between the predictions from the two 
extreme composition models, experimental data
do not provide a clean signature in favour of a given 
composition model. 

Another comparison between the TRD data and the Monte Carlo
predictions can be done by looking at the spectral
indices. For each  composition model
we fitted the underground energy spectra 
of single and double muons with the formula~\ref{eq:undspec},
assuming the same values of $\epsilon$ and $\beta$
as in sec.~\ref{sec:fit} and we compared the fit 
results with the TRD data. Table~\ref{tab:hemasfit}
shows the comparison of the data with the Monte Carlo 
predictions. As in the previous case, the 
single and double muon spectral indices do not allow 
to perform a study of the cosmic ray composition.

The sensitivity of our measurement to the primary cosmic 
ray composition is mainly limited by its precision and 
by the poor statistics of the high multiplicity muon events, 
that does not allow to reconstruct their spectra.
A detector with a larger area than the MACRO TRD and 
with a reduced systematics could enhance the precision
of the measurement of the single and double muon energies
and also allow to measure the energies of high multiplicity 
muons, that are more sensitive to the cosmic ray composition
in the energy region of the knee.


\section{Conclusions}

The MACRO TRD allowed the measurement of the energies
of muons penetrating in the Gran Sasso underground laboratories,
in the standard rock depth range from $3000$ to $6500 \units{hg/cm^{2}}$. 
For reconstructing the muon energy spectra 
we used an unfolding method and a best fit procedure. 
The average energies of single and double muons, 
evaluated with the unfolding technique described in 
sec.~\ref{sec:unfo}, are 
$\langle E \rangle_{1} = 270 \pm 3 (stat) \pm 18 (syst) \units{GeV}$ 
and $\langle E \rangle_{2} = 381 \pm 13 (stat) \pm 21 (syst)
\units{GeV}$.
The spectral indices for the approximate parameterization 
of the muon energy spectra (eq.~\ref{eq:undspec}), 
evaluated with the
best fit procedure described in sec.~\ref{sec:fit} are
$\alpha_{1} = 3.79 \pm 0.02 (stat) \pm 0.11 (syst)$
and $\alpha_{2} = 3.25 \pm 0.06 (stat) \pm 0.07 (syst)$.

We also performed a Monte Carlo simulation to study how the muon
energy spectra depend on the primary cosmic ray energy spectra
and composition, on their interactions with the atmosphere and
on the muon propagation in the rock.
Our data show that double muons are more energetic than
single ones in the standard rock depth range from $3000$ to 
$6500~hg/cm^{2}$, as predicted by our Monte Carlo simulation. 
The Monte Carlo simulation also showed that a measurement of the
energies of underground high multiplicity muons could provide useful
information about the primary cosmic ray composition. 
In fact, we remarked that the differences between the 
predictions from the various composition models become
relevant at high muon multiplicities, where the contribution
from heavy primaries is more significant.   

The energy spectra of single and double muons reconstructed 
from our data are in agreement with the Monte 
Carlo predictions obtained assuming two extreme cosmic ray 
composition models. This measurement do not allow to perform 
a cosmic ray composition study because the errors are
compatible with the gaps between the predictions 
from the two models. However, a measurement of the 
energy spectra of high multiplicity underground muons 
could provide a useful tool for investigating the primary 
cosmic ray composition in the energy region above the 
knee of the spectrum.


\bibliographystyle{unsrt}



\newpage

\begin{table}
\begin{center}
\begin{tabular}{||l|c|c|c|c||}
\hline
\hline
        & \bf{Module 1} & \bf{Module 2} & \bf{Module 3} & \bf{TRD} \\
\hline
\hline
\bf{Analyzed runs}    & 6134    & 3261    & 4218    &         \\
\bf{Bad runs}     & 3538    & 2160    & 1897    &         \\ 
\bf{Live time}  & 29004 \units{h} & 14942 \units{h} & 
18122 \units{h} & 62069 \units{h} \\
\hline
\bf{Single $\mu$ after cuts} & 
128903          & 66675         & 54712         & 250290        \\
\hline
\bf{Double $\mu$ after cuts} & 
8904            & 4576          & 4463          & 17943         \\
\hline
\hline
\end{tabular}
\end{center}
\caption{Summary of the data collected by the MACRO TRD. Third module
data refers only to the part of the module not equipped with
aluminum foils.}
\label{tab:summary}
\end{table}

\newpage

\begin{table}
\begin{center}
\begin{tabular}{|l|c|c||c|c||}
\hline
\hline
  & \multicolumn{2}{|c||}{\bf TRD DATA} & 
\multicolumn{2}{c||}{\bf MC SIMULATION} \\
\cline{2-5}
  & Number of & Average & Number of & Average \\
  & events & number of hits & events & number of hits \\  
\hline
\hline
\bf{Module 1} & $223$ & $2.06 \pm 0.12$ & $3273$ & $2.12 \pm 0.03$ \\
\bf{Modules 2 + 3} & $283$ & $2.05 \pm 0.13$ & $3273$ & $2.07 \pm 0.04$ \\
\hline
\hline
\end{tabular}
\end{center}
\caption{Stopping muon data: the measured hit distributions 
are compared with the predictions from the simulation.}
\label{tab:mustop}
\end{table}

\newpage

\begin{table}[h]
\begin{center}
\begin{tabular}{|l|c|c|c|}
\hline
\hline
   & $\mathbf{\langle E \rangle_{E< E_{0}}}$ & $\mathbf{f}$ & 
   $\mathbf{\langle E \rangle}$ \\
   & $\mathbf{(GeV)}$ & $\mathbf{\%}$ &   $\mathbf{(GeV)}$ \\
\hline
\bf{Single muons} & $195 \pm 2_{sta} \pm 15_{sys}$ & 
$4.5 \pm 0.1_{sta}\pm 0.7_{sys}$  & $270 \pm 3_{sta} \pm 18_{sys}$    \\
\bf{Double muons} & $234 \pm 11_{sta}\pm 18_{sys}$ & 
$9.0 \pm 0.5_{sta} \pm 1.0_{sys}$ & $381 \pm 13_{sta} \pm 21_{sys}$ \\
\hline
\hline
\end{tabular}
\end{center}
\caption{Average energy of underground single and double muons, 
with residual energy below the saturation threshold ($E_{0}=1\ TeV$) 
of our TRD (column 2); fraction of events with energy above 
the threshold (column 3); average energy of all events below and 
above threshold (column 4).}
\label{tab:energy}
\end{table}

\newpage

\begin{table}[!b]
\begin{center}
\begin{tabular}{||l|c|c|c|c||}
\multicolumn{5}{c}{\bf{Light model}} \\
\hline
Group   & $K_{1} (\units{m^{-2}s^{-1}sr^{-1}GeV^{\gamma_{1}}})$ &
$\gamma_{1}$ & $E_{cut}(\units{GeV})$ & $\gamma_{2}$ \\
\hline
p       & $1.5 \times 10^4$     & 1.71  &       $2.0 \times 10^4$ &     \\
        & $1.87 \times 10^3$    & 1.50  &       $3.0 \times 10^6$ & 2.0 \\
He      & $5.69 \times 10^3$    & 1.71  &       $3.0 \times 10^6$ & 2.0 \\
CNO     & $3.30 \times 10^3$    & 1.71  &       $3.0 \times 10^6$ & 2.0 \\
Mg      & $2.60 \times 10^3$    & 1.71  &       $3.0 \times 10^6$ & 2.0 \\
Fe      & $3.48 \times 10^3$    & 1.71  &       $3.0 \times 10^6$ & 2.0 \\
\hline
\end{tabular}
\vspace{0.5cm}
\begin{tabular}{||l|c|c|c|c||}
\multicolumn{5}{c}{\bf{Heavy model}} \\
\hline
Group   & $K_{1} (\units{m^{-2}s^{-1}sr^{-1}GeV^{\gamma_{1}}})$ &
$\gamma_{1}$ & $E_{cut}(\units{GeV})$ & $\gamma_{2}$ \\
\hline
p       & $1.5 \times 10^4$     & 1.71  &       $1.0 \times 10^5$ & 2.0 \\
He      & $5.69 \times 10^3$    & 1.71  &       $2.0 \times 10^5$ & 2.0 \\
CNO     & $3.30 \times 10^3$    & 1.71  &       $7.0 \times 10^5$ & 2.0 \\
Mg      & $2.60 \times 10^3$    & 1.71  &       $1.2 \times 10^6$ & 2.0 \\
Fe      & $3.48 \times 10^3$    & 1.36  &       $2.7 \times 10^6$ & 2.0 \\
\hline
\end{tabular}
\end{center}
\caption{Parameters of the primary cosmic ray energy spectra 
according the ``light'' and ``heavy'' composition models.}
\label{tab:models}
\end{table}

\newpage

\begin{table}[!b]
\begin{center}
\begin{tabular}{||l||c|c||c|c||}
\hline
\hline
  & \multicolumn{2}{|c||}{\bf Single muons} & 
\multicolumn{2}{c||}{\bf Double muons} \\
\cline{2-5}
  & $\alpha$ & $\chi^{2} /d.o.f.$ & $\alpha$ & $\chi^{2} /d.o.f.$ \\
\hline
{\bf TRD data} & $3.79 \pm 0.02 \pm 0.11$ & $1.51$ & 
$3.25 \pm 0.06 \pm 0.07$ & $0.55$ \\
\hline 
Light model & $3.70$ & $1.55$ & $3.14$ & $0.92$ \\
Heavy model & $3.84$ & $1.73$ & $3.36$ & $1.02$ \\
\hline
\hline
\end{tabular}
\end{center}
\caption{Results of the fits of underground muon energy spectra 
with the formula~\ref{eq:undspec}. TRD data (with the statistical
and systematic error) are compared with the
Monte Carlo predictions obtained assuming two different 
composition models for primary cosmic rays.} 
\label{tab:hemasfit}
\end{table}


\newpage

\newpage

\newpage

\newpage

\newpage

\newpage

\newpage

\newpage

\newpage

\end{document}